\documentclass[11pt,titlepage]{article}
\usepackage[T1]{fontenc}
\usepackage[latin9]{inputenc}
\usepackage{amsthm}
\usepackage{amsmath}
\usepackage{graphicx}
\usepackage[round,numbers,sort&compress]{natbib} 
\usepackage{geometry}
\makeatletter
\@ifundefined{textcolor}{}
{%
 \definecolor{BLACK}{gray}{0}
 \definecolor{WHITE}{gray}{1}
 \definecolor{RED}{rgb}{1,0,0}
 \definecolor{GREEN}{rgb}{0,1,0}
 \definecolor{BLUE}{rgb}{0,0,1}
 \definecolor{CYAN}{cmyk}{1,0,0,0}
 \definecolor{MAGENTA}{cmyk}{0,1,0,0}
 \definecolor{YELLOW}{cmyk}{0,0,1,0}
 }

\makeatletter

\newcommand{\Rmnum}[1]{\expandafter\@slowromancap\romannumeral #1@}
\makeatother

\newcommand{\be}{\begin{equation}}
\newcommand{\ee}{\end{equation}}

\newcommand{\avg}[1]{\langle #1 \rangle}

\def\lsim{\mathrel{\rlap{\lower4pt\hbox{$\sim$}}
    \raise1pt\hbox{$<$}}}                
\def\gsim{\mathrel{\rlap{\lower4pt\hbox{$\sim$}}
    \raise1pt\hbox{$>$}}}             

\newcommand{\beginsupplement}{%
        \setcounter{section}{0}
        \renewcommand{\thesection}{S\arabic{section}}%
        \setcounter{table}{0}
        \renewcommand{\thetable}{S\arabic{table}}%
        \setcounter{figure}{0}
        \renewcommand{\thefigure}{S\arabic{figure}}%
     }

\renewcommand\[{\begin{equation}}
\renewcommand\]{\end{equation}}
\usepackage[english]{babel}
\addto\captionsenglish{}
\makeatother

\usepackage{babel}

\begin{document}

\title{Determination of single-bond association kinetics by tethered particle motion: concept and simulations.}
\author{Koen E. Merkus \\
	Department of Applied Physics, \\
	  Eindhoven University of Technology, Eindhoven, The Netherlands\\
    \and Menno W.J. Prins\thanks{
           Corresponding author.  Address: 
Department of Applied Physics
Eindhoven University of Technology,
NL-5600 MB Eindhoven, The Netherlands} \\
    Department of Applied Physics, \\
    Department of Biomedical Engineering,  \\
    Institute for Complex Molecular Systems, \\
	  Eindhoven University of Technology, Eindhoven, The Netherlands
	\and Cornelis Storm \\
	Department of Applied Physics and\\
	Institute for Complex Molecular Systems \\
	  Eindhoven University of Technology, Eindhoven, The Netherlands}

\date{\today}

\pagestyle{myheadings}
\markright{TPM association kinetics}

\maketitle


\abstract{
Tethered particle motion (TPM) --- the motion of a micro- or nanoparticle tethered to a substrate by a
macromolecule --- is a system that has proven extremely useful for its ability to reveal physical features of the tether, because the thermal motion of the bound particle reports sensitively on parameters like the length, the rigidity, or the folding state of its tether. In this paper, we discuss a novel application of TPM, surveying its utility in probing the kinetics of single {\em secondary} bonds: bonds that form and break between the tethered particle and a substrate due, for instance, to receptor/ligand pairs on particle and substrate. Much like the tether itself affects the motion pattern, so do the presence and absence of such secondary connections. Keeping the tether properties constant, we demonstrate how raw positional TPM data may be parsed to generate detailed insights into the association and dissociation kinetics of single secondary bonds. We do this using coarse-grained molecular dynamics simulations, specifically developed to treat the motion of particles close to interfaces. 

\emph{Keywords:} tethered particle motion, bond kinetics, molecular dynamics.}
\clearpage

\section{Introduction}
The precise measurement of the binding and unbinding kinetics of single biological bonds has been the ambition of an active and rapidly developing field for over two decades now~\cite{Robert2007}. A multitude of experimental methods to probe bonds at the single-molecule level are currently available, and include AFM~\cite{florin1994adhesion}, optical tweezers~\cite{thoumine2000short}, magnetic tweezers~\cite{danilowicz2005dissociation}, \cite{jacob2012quantification}, laminar flow chambers~\cite{kaplanski1993granulocyte}, total internal reflection fluorescence microscopy~\cite{schneckenburger2005total}, \cite{jungmann2010single}, interferometric imaging~\cite{piliarik2014direct}, plasmonic sensing~\cite{beuwer2015stochastic} and acoustic force spectroscopy~\cite{sitters2015acoustic}. Moreover, the focus on further development of the single-molecule toolbox is projected to intensify in the direct future~\cite{VanOijen2011}. In this paper, we direct our attention to a single-molecule property for which currently few tools are available: the association kinetics of a single pair of noncovalently bonding molecules. A first complicating factor in measuring these properties is the generic difficulty to disentangle extrinsic and intrinsic factors: association is a strongly distance-dependent process that, trivially, may only occur when the two binding partners are within touching proximity. A second complicating factor is that molecular association is a stochastic process, so one needs to be able to gather statistical data of repeated events in order to be able to extract association parameters with sufficient precision. This requires a stable molecular system and a readout arrangement suited for long observation times.
\\[2mm]
Here we propose a method based on Tethered Particle Motion (TPM) to address the abovementioned problems. The method allows for separation of the encounter and association processes, it allows for repeated probing of the same system and it allows for long observation times. The method is based on measuring bond kinetics by tracking the motion of a molecularly tethered particle that can form secondary bonds with a substrate, as sketched in Fig~\ref{fig:fig1}. In its rawest form, the data collected in such an experiment consist of a time series $\vec R(t_i)$ of the particle position projected onto the substrate. We will call such a trace a {\em motion pattern}. The basic idea is straightforward: if a secondary bond is present, the motion pattern of the bead will change as it is now confined by two, rather than one, bonds to the substrate~\cite{Visser}. If no secondary bond is present, the motion pattern is that of a regular noninteracting TPM system. Thus, the motion pattern itself reports on the binding state. There is, however, more information in the time-varying motion pattern: the dynamics of the {\em switching} between the different (bound and free) motion patters reports on the kinetics of the secondary bond.
\\[2mm]
In what follows, we present first the basic concepts and definitions required to measure association kinetics by TPM, and then present Molecular Dynamics (MD) simulations of the process. The insight from these simulations permit us to devise a protocol to extract the association rates from raw experimental data. We provide proof-of-principle for our method, validating the protocol using simulated raw data.
\section{Motion of tethered particles with secondary bonds}\label{sec:motion}
Tethered particle motion is a proven tool in biophysics. It has been used to determine the transcription of RNA polymerase~\cite{DorothyA.SchaferJeffGellesMichaelP.Sheetz1991}, the persistence length of DNA~\cite{Brinkers2009} and the looping kinetics of DNA~\cite{milstein2011bead}. In all previous TPM-based research, the focus has been on properties and interactions of the tether, rather than those of the particle. For the most part, the particle has been a large (and therefore easily visualized) marker for the end of the tether.

There is, however, no reason why one should not assign further functionality to the particle itself. In particular, it will prove useful to consider particles that may form additional bonds with the substrate, besides the one effected by the tether. In what follows, we will assume the dsDNA tether not to change, and use it solely as a means to keep the functionalized particle close to the substrate and confine its random thermal motion. As explained, there is useful information in both the instantaneous magnitude and the time-dependence of the raw signal $\vec R(t_i)$. However, while all this information is undoubtedly present, a crucial question is whether $\vec R(t_i)$ may be dissected to isolate the association rate. We show that this is indeed possible.

A clear advantage of our method is that it is easily parallelized: multiple particles can be tethered to the same substrate, the measurement of their trajectories may be done robustly in wide field optics, and the change in motion pattern of even a single particle due to binding may constitute a detection and permit the determination of rates --- multiple detections obviously improving the accuracy of the method.\\[2mm]
Fig.~\ref{fig:fig1}a presents the basic concept of TPM-based single molecule measurements that we propose. The particle is in one of three states: free, encounter, or bound. The {\em free} state is one in which the particle is only bound to the substrate by the tether. The {\em encounter} state is a conformation in which the particle and the substrate are close enough for a secondary bond to form, but this bond is not actually connected. The {\em bound} state, finally, is where the secondary bond is formed. Four distinct rates characterize the transitions back and forth between the three particle states. By observing the motion pattern we may distinguish only between states where the secondary bond is not present (free, encounter) and the one where it is (bound). The entangled, distance-dependent character of this process is obvious: The transition from the free to the bound state is a two-step process, that must necessarily pass through the encounter state. An experiment where no changes in binding pattern are observed may either have a complexation rate $k_{\rm c}$ that is low with respect to the reciprocal timescale of the experiment, a dissociation rate $k_{\rm off}$ that is high with respect to the reciprocal time resolution of the experimental detection method, or an encounter rate $k_{\rm enc}$ that is so low that no opportunity for binding occurs on the timescale of the experiment. The first two relate to an intrinsic property of the single bond, the third relates to an extrinsic effect, dependent on the geometry of the particle-tether construct and its Brownian motion. Clearly it is important to separate the intrinsic and extrinsic effects. The point of this paper is to demonstrate that the contribution of the particle's Brownian motion to the overall motion can be modeled and understood using MD simulations. Applying this principle allows otherwise inaccessible single-bond properties to be extracted from raw experimental data.


\section{TPM and changing motion patterns}

To be able to put real numbers on our axes, and to remain close to experimentally feasible dimensions we will consider for the most part a 50 nm double-stranded DNA (dsDNA) tether that attaches a 1 $\mu$m diameter particle to a substrate~\cite{Visser}. In typical TPM systems the in-plane motion of this particle is tracked over time, so that a two-dimensional projection $\vec R(t_i)$ of the movement of the particles is obtained~\cite{milstein2011bead},~\cite{Fan2012}, \cite{kumar2014enhanced}. When such a particle is repeatedly imaged within several consecutive time intervals $\Delta t=t_{i+1}-t_i$, the combined result is a motion pattern like the one shown in the left panel of Fig.~\ref{fig:fig1}b.

We consider now the case where both the particle and the substrate are coated with complementary binding molecules. This results in the possibility of a tethered particle to form the secondary bond with the substrate. When such a secondary bond is formed, the motion pattern is significantly altered: It is no longer axisymmetric, and much more localized as shown in the right panel of Fig.~\ref{fig:fig1}b.\\[2mm]

The in-plane distance that the particle travels between two frames is indicative of the presence of a secondary bond. We will refer to this traveled distance as the {\em step size}, defined as $R_{step}(t) = |\vec{R}(t_{i+1}) - \vec{R}(t_i)|$. Since a secondary bond results in additional confinement of the motion of our tethered particle, the average step size is reduced by secondary bonds. This is graphically represented in the distribution of step sizes in Fig.~\ref{fig:fig1}. Moreover, the shape of the motion pattern in the bound state is the result of the specific position of the binding molecules. The typical time between formation and dissociation of these secondary bonds is the kinetic data that we aim to relate to the kinetic binding properties of the individual molecules that form the bond.\\[2mm]
A more detailed description on how the step size may be used to distinguish between states can be found in~\cite{Max}. Alternatively, one might consider the Brownian motion amplitude as indicator of the formation of a secondary bond, as was previously used to study the looping state of DNA~\cite{Fan2012}.\\[2mm]
The step size itself is a potential reporter for the presence or absence of a secondary bond. There is, however, additional information in the switches in magnitude of the step size, and this additional information has not been tapped into yet. To extract it, we must  understand the temporal behavior of this system. This is why we now turn to molecular dynamics (MD) simulations.


\section{Simulation methods} 

We perform Langevin dynamics simulations using LAMMPS molecular dynamics package~\cite{plimpton1995fast} with one spatially extended spherical particle, the TPM particle, that is connected to a flexible tether, which is represented by a string of $N$ point particles in a bead-spring model. Our simulation is thus a coarse-grained MD simulation to describe the motion of a TPM system.
\subsection{MD algorithm}
In the Langevin dynamics method, each particle is subject to conservative, drag and random forces, $\vec F_c$, $\vec F_d$ and $\vec F_r$,  respectively, and obeys the following translational equation of motion:
\begin{equation}
M \ddot{\vec r} = \vec F_c + \vec F_d + \vec F_r, \label{eq:sum_forces}
\end{equation}
where $M$ is the particle's mass and $\ddot{\vec r}$ is the particle's acceleration. Excluded volume, bonding and angle-bend interactions are explicitly included in $\vec F_c$, as described in section~\ref{sec:interactions}. The drag force is given by $\vec F_d = - \gamma \dot{\vec r}$, where $\gamma$ is the drag coefficient and $\dot{\vec r}$ is the particle's velocity. The drag coefficient is described in more detail in section~\ref{sec:drag}. The drag force $\vec F_d$ and the random force $\vec F_r$ are both the result of the interaction with the solvent and, by extension, these forces are related. In particular, the fluctuation-dissipation theorem tells us that~\cite{Zhang2003}
\begin{equation}
\avg{\vec F_r(t) \cdot \vec F_r(t')} = 6 k_B T \gamma \delta(t-t').\label{eq:flucdis},
\end{equation}
with $k_B$ Boltzmann's constant and $T$ the temperature.\\[2mm]
For a spatially extended particle the rotational motion has to be taken into account as well. This results in the rotational counterpart of equation~\ref{eq:sum_forces}:
\begin{equation}
I \ddot{\vec \phi} = \vec \tau_c + \vec \tau_d + \vec \tau_r,
\end{equation}
where $I$ denotes the moment of inertia, $\ddot{\vec \phi}$ is the angular acceleration and the conservative, drag and random torque are given by $\vec \tau_c$, $\vec \tau_d$ and $\vec \tau_r$ respectively. Similar to the translational force, the translational drag is given by $\vec \tau_d =  - \gamma_{rot} \dot{\vec \phi}$, where $\gamma_{rot}$ is the rotational drag coefficient and $\dot{\vec \phi}$ is the angular velocity. Once more, the fluctuation-dissipation theorem provides us a relation for the effects caused by the solvent, \emph{i.e.,}
\begin{equation}
\avg{\vec \tau_r(t) \cdot \vec \tau_r(t')} = 6 k_B T \gamma_{rot} \delta(t-t').
\end{equation}

\subsection{Interactions}\label{sec:interactions}
There are two main types of interactions included in our simulations: firstly, interactions that prescribe the behavior of the tether and secondly, steric exclusion effects.\\[2mm]
A bead-spring model is used to represent the tether in the simulations. These beads are held together by a harmonic bond potential given by
\begin{equation}
U_{bond} = K_b(r-r_0)^2, \label{eq:bond_potential}
\end{equation}
with bond coefficient $K_b$, $r_0$ the rest distance between two beads and $r$ the distance between two beads. To ensure that bond lengths are essentially fixed during the simulations, we choose a large value for the strength of the bond potential, $K_b = 50 k_B T/r_0^2$~\cite{Naderi2014}, where $k_B T$ is the thermal energy.
\\[2mm]
To include the limited flexibility of a typical dsDNA tether, we include an angle-bending potential
\begin{equation}
U_{angle} = K_a \theta^2
\end{equation}
where $K_a$ is the angle-bending coefficient and $\theta$ is the angle between two adjacent springs. In the limit of $r_0/l_p \rightarrow 0$, the angle-bending coefficient can be related to the thermal energy $k_B T$, the persistence length $l_p$ and the rest distance $r_0$ by $K_a= \frac{k_B T l_p}{2 r_0}$~\cite{Underhill2004}. We construct our model so that $r_0/l_p \ll 1$, which justifies the use of this angle-bend potential.\\[2mm]
Moreover, three relevant steric exclusion mechanisms are present in a TPM system: tether-substrate exclusion, particle-substrate exclusion, and tether-particle exclusion. For the tether-substrate exclusion we use the repulsive part of the Lennard-Jones potential
\begin{equation}
U_{LJ}(r)  = 4 \epsilon \left[ \left( \frac{\sigma}{r}\right)^{12} - \left(\frac{\sigma}{r}\right)^6\right], \quad \; \; r < r_c\label{eq:LJ}
\end{equation}
where the energy $\epsilon$ and distance $\sigma$ are the parameters that define the potential and $r_c=2^{1/6}\sigma$ so that only the repulsive part is used. $r$ describes the distance beween the interacting elements.\\[2mm]
To ensure a steep potential at the edge of the particle, an extra parameter is required. Therefore, for the interactions that involve the particle we use
\begin{equation}
U_{LJ,exp}(r)  = 4 \epsilon \left[ \left( \frac{\sigma}{r-\Delta}\right)^{12} - \left(\frac{\sigma}{r-\Delta}\right)^6\right], \quad \; \; r < r_c + \Delta \label{eq:LJ_expand},
\end{equation}
as the expanded potential with extra parameter $\Delta$. \\[2mm]
For $\epsilon$ we have used a value of 100 $k_B T$ to ensure repulsion and minimize the overlap of particles. For $\sigma$ we have chosen 1 nm, which is small enough to reproduce geometrically sensible results, but large enough to create a steepness of the potential that is computationally acceptable. To model a particle with radius 500 nm, we have chosen $\Delta = 499$ nm.

\subsection{Drag coefficients}\label{sec:drag}
\subsubsection{Drag on particle}\label{sec:drag_bead}
An important feature of a relatively big particle on a small tether is that it is at all times close to the substrate, \emph{i.e.,} the distance between the substrate and the edge of the particle is much smaller than the radius of the particle. Hydrodynamic wall effects are therefore important~\cite{Brenner1961},~\cite{happel1983low}, \emph{i.e.,} the drag on a sphere is no longer given by the isotropic Stokes drag, but the drag coefficient is elevated and no longer isotropic. \\[2mm]
For a spherical particle with radius $R$ and distance $z$ between the particle center and the substrate, the parallel and perpendicular drag coefficients are given by~\cite{Scha2007}
\begin{align}
\gamma_{\parallel} &= \frac{\gamma_0}{1-\frac{9}{16}q + \frac{1}{8}q^3-\frac{45}{256}q^4-\frac{1}{16}q^5},\\
\gamma_{\bot} &=  \frac{\gamma_0}{1-\frac{9}{8}q + \frac{1}{2}q-\frac{57}{100}q^4+\frac{1}{5}q^5},
\end{align}
where $q=R/z$, $\eta$ is the dynamic viscosity of the solvent and $\gamma_0 = 6 \pi \eta R$, the Stokes drag in an unbounded liquid.

\subsubsection{Drag on tether}
The tether is modeled by a bead-spring system consisting of $N_{beads}$ beads, where every bead is subject to a Stokes drag force with drag coefficient
\begin{equation}
\gamma_{teth} = 6 \pi \eta R_{hy},
\end{equation}
where $R_{hy}$ is the so-called hydrodynamic radius of the beads: an effective parameter that determines the amount of drag on the tether.\\[2mm]
In general, a polymer in solution experiences  different drag in the direction parallel and perpendicular to its axis. In a TPM experiment, the ends of the tether are attached to the bead and the substrate, so that the tether predominantly moves in the direction perpendicular to its axis. Due to the fact that the drag on the particle, more than the drag on the tether, provides the dominant contribution to the observable timescales in the system and due to considerations of computational efficiency, we have approximated the drag on the tether by the drag on a cylinder in the perpendicular direction~\cite{doi1986theory}, distributed evenly across all beads without preferential direction. This leads to
\begin{equation}
\gamma_{teth} = \frac{4 \pi\eta l}{N_{beads} \ln(l/b)},
\end{equation}
where $l$ is the contour length of the tether, $b$ is the width of the cylinder and for dsDNA we use $b=2$ nm~\cite{phillips2012physical}. Since the hydrodynamic radius $R_{hy}$ is typically much smaller than the average distance of the beads to the substrate and those beads that are close to the surface move very little, we neglect the hydrodynamic surface effects on the tether and apply a homogeneous Stokes drag to each bead.

\subsection{Validation of the spatial encounter distribution and diffusion kinetics}
We now compare spatial and kinetic results from our MD simulations to results from reference methods. First, we compare our simulation results to results from previously developed Monte Carlo (MC) simulations,  which yield the probability distribution of particle positions in the equilibrium state~\cite{Visser}. The spatial encounter distribution describes the probability as function of the in-plane position that a particle is within the encounter interaction range of the substrate.  In Fig.~\ref{fig:benchmark}a the spatial encounter distribution is compared for the MC results and for MD results with a varying number of beads that make up the tether. We observe that for an increasing number of beads the two distributions converge. Based on this comparison, we have used a tether consisting of 10 beads in our MD simulations.
\\[2mm]
Second, we compare results from our MD simulations to the analytical expression for Brownian motion. From conventional Brownian motion theory, it is known that the in-plane mean-squared distance traveled by a free diffusing particle is given by
\begin{equation}
\langle R_{\parallel}^2 \rangle =  4 D t
\end{equation}
as opposed to the factor 6 for 3D. Here $D$ stands for the diffusion coefficient and $t$ for time.\\[2mm]
In the TPM system, the drag on the particle is increased near the surface, as described in section~\ref{sec:drag_bead}. This results in a lower effective diffusion coefficient $D_{\parallel}$. We have performed 3000 MD simulations starting out with a TPM system in upright position and we plot the simulated $\langle R_{\parallel}^2 \rangle$ and the calculated $4 D_{\parallel} t$ in one figure. The result are presented in Fig.~\ref{fig:benchmark}b.\\[2mm]
For small times (see the inset) the MD simulation results indeed correspond to the analytical relation for a free particle. For larger times differences occur, caused by the tether that pulls the particle on average closer to the substrate, which results in a higher drag and a lower diffusivity.


\section{Determining association kinetics}

\subsection{General algorithm}\label{sec:algorithm}
Now that we have introduced the relevant concepts and properties that are involved in the measurement of single-bond kinetics using TPM, we outline an algorithm that allows one to process and interpret measurement data. A schematic outline of the algorithm can be found in Fig.~\ref{fig:scheme}.
We will further expand on either of these steps in this section and we will clarify how these steps enable the extraction of single-bond data from a TPM experiment.
\\[2mm]
The starting point of our algorithm is the data of a TPM experiment. This is generally the 2-D projected position of the particle captured multiple times in a sequence of time intervals.
The initial analysis of this data aims to isolate the captured positions that correspond to the particle being in the bound state. A means of doing this is by reviewing the step size: the in-plane distance the particle travels between frames. By averaging this over multiple frames and using two separate thresholds, as described in~\cite{Max}, individual binding events can be isolated from the experimental data, an example of this can be found in Fig~\ref{fig:analysis}a.
\\[2mm]
Having extracted the frames that correspond to the bound state, these frames are put together to compose the time-independent bound motion pattern. More sophisticated metrics may be employed, but for now we characterize the bound motion pattern using three parameters: its length $L$, its width $W$ and its average distance to the central axis $D$. These parameters are indicated by arrows in Fig.~\ref{fig:analysis}b.

\vspace{2mm}
\noindent
With the bound motion patterns thus characterized, the next step is to determine the location of the binding molecules that is most likely to correspond to that particular binding event. This is where simulation data comes in: using the results from simulations with varying binding spots we can compile the functional relation between binding spot and motion pattern, allowing us to translate the values $(L,W,D)$ into values for $d_s$ and $d_p$, the distances along substrate and particle where binding most likely occurred. Thus, our simulations permit us to extract new information from experimental data.\\[2mm]
Subsequently, simulations are used once more to determine the probability $P_{\rm enc}$ that these two binding spots are within the interaction range. We consider here the case where only one molecule is present on the particle and only one on the substrate, but our method is easily extended to include different coverages. Whether this is required of course depends on the specific experimental settings, and $P_{\rm enc}$ should be determined accordingly. Finally, by acquiring the distribution of times between binding events and factoring out $P_{\rm enc}$ we can isolate the molecular complexation rate $k_{\rm c}$. If the apparent binding rate is given by $\kappa$, then we can find $k_{\rm c}$ using the relation
\begin{equation}
\kappa = P_{\rm enc} \, k_{\rm c}.
\end{equation}
\\[2mm]
In summary, by using simulation results we have created the possibility to extract otherwise inaccessible single-molecule data from TPM experiments. 

\subsection{Example application}
As a proof-of-principle and as a means of demonstrating our algorithm we have constructed a mock experiment to generate pseudo-TPM data. This model requires three geometric parameters and four rates as input. The geometric parameters $L$, $W$ and $D$ describe the shape of the bound motion pattern. It uses preset values for the rates $k_{\rm enc}$, $k_{\rm sep}$, $k_{\rm c}$ and $k_{\rm off}$, and generates a motion pattern in time as output. The challenge to our analysis protocol is now to back out the value of $k_{\rm c}$ in the manner described above. The input values are listed in table~\ref{tab:mock}.
\\[2mm]
The mock model steps through time, and at every time point the system is in either the `free', `encounter' or `bound' state. Every step the system may change state and this happens with a probability determined by the relevant transition rates. \\[2mm]
In the simulations of the system with a secondary bond, the binding molecule had a Y-shape and a fully-stretched length of 15 nm. In the simulations without secondary bond we considered this 15 nm as the encounter range $d_{\rm enc}$.\\[2mm]
In the mock model, every time step a $X$- and $Y$-coordinate of the particle are generated. When the system is in the `free'-state, a random point in a circle with a radius of 220 nm is generated, \emph{i.e.,} within the typical radius of `free' motion patterns in our system. When the system is not in the free state, and thus in the `encounter'- or `bound'-state, a random point within a confining ellipse, with given $L$, $W$ and $D$, is generated. Clearly this results in points that are, on average, closer to each other in the bound state than in any of the unbound states. This can be seen in Fig.~\ref{fig:analysis}a, where the step size obtained from our mock model is represented. The data points are generated with a frame rate of 30 Hz, so the step size corresponds to the distance traveled in 0.033 s. An averaging window of 30 frames is used, so the black line corresponds to the distance traveled in 1 s.
\\[2mm]
Using the obtained step size and averaged step size, we now isolate the points of the motion pattern that are classified as belonging to the `bound' state so that we end up with a motion pattern similar to the pattern in Fig.~\ref{fig:analysis}b. As is indicated in this figure, we can extract the $L$, $W$ and $D$ from this figure. This leads to a value of $L=247$ nm, $W=141$ nm and $D=150$ nm, leading to the most probable binding spots $d_{p} = 160$ nm and $d_{s} = 200$ nm.\\[2mm]
Using simulation results of a free tethered particle we know that the probability of a binder on the substrate and on the particle being within interaction range $d_{\rm enc}$ is $P_{\rm enc} = 1.2 \cdot 10^{-4}$. This result is obtained by dividing the average number of frames in a simulation in which the two binders are within interaction range through the total number of frames in a simulation. We measure the time it takes for a free particle to become bound throughout the whole experiment to come up with a Cumulative Distribution Function (CDF). This distribution is fit to an exponential function, and the result is represented in Fig.~\ref{fig:analysis}c. From the fit we obtain the fitting parameter $\kappa=(1.7 \pm 0.3) \cdot 10^{-3}$ s$^{-1}$ (mean $\pm$ standard deviation).\\[2mm]
Factoring out the encounter probability $P_{\rm enc}$, we end up with a value of $k_{\rm c} = (1.4 \pm 0.2) \cdot 10^1$ s$^{-1}$, which is to be compared to the input value $k_{\rm c} = 1.7 \cdot 10^1$ s$^{-1}$ that was not used to generate the data. We hypothesize that the underestimation compared to the actual value may be attributed to a structural issue, also present in experiments, which is, that some events are too short lived to be resolved (shorter than one time step) and thus there is a systematic underestimation of both the bound time and the association rate. An averaging window to detect bonds of 1 s has been used, while the typical dissociation time is 10 s ($1/k_{\rm off}$). Therefore, a detectable difference between the apparent association rate and the actual association rate is expected. In further refinements, this too may be corrected for using the simulations.

\begin{table}[]
\centering
\caption{Input values for the mock model}\label{tab:mock}
\label{my-label}
\begin{tabular}{c|c|c}
parameter & value & units \\
\hline
$k_{\rm enc}$ & 1.0 & s$^{-1}$  \\
$k_{\rm sep}$ & 8.3$\cdot 10^3$ & s$^{-1}$\\
$k_{\rm c} $ & 1.7 $\cdot 10^1$ & s$^{-1}$\\
$k_{\rm off}$ & 0.1 & s$^{-1}$\\
$L$ & 247 & nm\\
$W$ & 141 & nm\\
$D$ & 150 & nm\\
\end{tabular}
\end{table}


\section{Conclusions and Outlook}
We have shown that by understanding the contribution of Brownian motion to the overall motion of a tethered particle, a TPM system allows for the probing of single bonds. We have focused on the basic principles of this novel approach. Several opportunities for future research arise. On the one hand, the understanding of the system and the corresponding simulations could be further developed. On the other hand, experimental validation of this method would be a crucial next cornerstone.\\[2mm]
The parameter that we only have a rough approximation for is the encounter distance $d_{\rm enc}$. The next step in developing a TPM based association measurement would be to increase our understanding of the encounter distance.
Two potential experiments come to mind. Firstly, one could investigate the influence of the encounter distance by using linker molecules with different lengths. The length of linker molecules between binder A and the substrate, and between binder B and the particle, will profoundly influence the encounter distance corresponding to the binder complex. If the apparent association rate $\kappa$ in an experiment with varying linker lengths $\ell$ would display comparable behavior to the encounter probability $P_{\rm enc}$ as function of the encounter distance $d_{\rm enc}$ in the simulations, that would be a strong experimental justification of our approach.
Secondly, by using several different binding molecules, a range of complexation rates $k_{\rm c}$ may be probed. In the upper limit, every encounter will lead to a bond, so for high complexation rates the apparent association rate should converge to $\kappa = P_{\rm enc}$.\\[2mm]
On the simulations side, the next step would be to incorporate the molecular complexation process. The complexation rate $k_{\rm c}$ is the result of distance-dependent physicochemical molecular interactions: charge interactions, van der Waals interactions, steric effects, hydrogen bonds, hydrophobic effects, etc. Simulations with a higher level of detail could provide more insight in the molecular processes involved in TPM experiments.\\[2mm]
In summary, what we show here is that the measurement and analysis of surface-binding data in TPM experiments provides a new window on single-molecule association dynamics, adding a novel modality to establish these important properties in experiments.\\
Given the new measurement concept proposed in this paper and the results from the simulations, we believe that the measurement and kinetic analysis of surface-binding data in TPM experiments holds promise to become a new modality for studies on single-molecule association kinetics.\\[2mm]


\section{Acknowledgments}
We thank Max Scheepers, Emiel Visser and Leo van IJzendoorn for numerous valuable discussions. We thank Emiel Visser for providing us the code for the MC simulations and Stefan Paquay for his continuous helpfulness in developing MD simulations.

\bibliography{AssociationKineticsTPM_references}

\begin{thebibliography}{29}
\providecommand{\url}[1]{\texttt{#1}}
\providecommand{\urlprefix}{ }

\bibitem[Robert et~al.(2007)Robert, Benoliel, Pierres, and
  Bongrand]{Robert2007}
Robert, P., A.-M. Benoliel, A.~Pierres, and P.~Bongrand, 2007.
\newblock What is the biological relevance of the specific bond properties
  revealed by single-molecule studies?
\newblock \emph{Journal of Molecular Recognition} 20:432--447.

\bibitem[Florin et~al.(1994)Florin, Moy, and Gaub]{florin1994adhesion}
Florin, E.-L., V.~T. Moy, and H.~E. Gaub, 1994.
\newblock Adhesion forces between individual ligand-receptor pairs.
\newblock \emph{Science} 264:415--417.

\bibitem[Thoumine et~al.(2000)Thoumine, Kocian, Kottelat, and
  Meister]{thoumine2000short}
Thoumine, O., P.~Kocian, A.~Kottelat, and J.-J. Meister, 2000.
\newblock Short-term binding of fibroblasts to fibronectin: optical tweezers
  experiments and probabilistic analysis.
\newblock \emph{European Biophysics Journal} 29:398--408.

\bibitem[Danilowicz et~al.(2005)Danilowicz, Greenfield, and
  Prentiss]{danilowicz2005dissociation}
Danilowicz, C., D.~Greenfield, and M.~Prentiss, 2005.
\newblock Dissociation of ligand-receptor complexes using magnetic tweezers.
\newblock \emph{Analytical chemistry} 77:3023--3028.

\bibitem[Jacob et~al.(2012)Jacob, van IJzendoorn, de~Jong, and
  Prins]{jacob2012quantification}
Jacob, A., L.~J. van IJzendoorn, A.~M. de~Jong, and M.~W. Prins, 2012.
\newblock Quantification of protein--ligand dissociation kinetics in
  heterogeneous affinity assays.
\newblock \emph{Analytical chemistry} 84:9287--9294.

\bibitem[Kaplanski et~al.(1993)Kaplanski, Farnarier, Tissot, Pierres, Benoliel,
  Alessi, Kaplanski, and Bongrand]{kaplanski1993granulocyte}
Kaplanski, G., C.~Farnarier, O.~Tissot, A.~Pierres, A.~Benoliel, M.~Alessi,
  S.~Kaplanski, and P.~Bongrand, 1993.
\newblock Granulocyte-endothelium initial adhesion. Analysis of transient
  binding events mediated by E-selectin in a laminar shear flow.
\newblock \emph{Biophysical journal} 64:1922.

\bibitem[Schneckenburger(2005)]{schneckenburger2005total}
Schneckenburger, H., 2005.
\newblock Total internal reflection fluorescence microscopy: technical
  innovations and novel applications.
\newblock \emph{Current opinion in biotechnology} 16:13--18.

\bibitem[Jungmann et~al.(2010)Jungmann, Steinhauer, Scheible, Kuzyk, Tinnefeld,
  and Simmel]{jungmann2010single}
Jungmann, R., C.~Steinhauer, M.~Scheible, A.~Kuzyk, P.~Tinnefeld, and F.~C.
  Simmel, 2010.
\newblock Single-molecule kinetics and super-resolution microscopy by
  fluorescence imaging of transient binding on DNA origami.
\newblock \emph{Nano letters} 10:4756--4761.

\bibitem[Piliarik and Sandoghdar(2014)]{piliarik2014direct}
Piliarik, M., and V.~Sandoghdar, 2014.
\newblock Direct optical sensing of single unlabelled proteins and
  super-resolution imaging of their binding sites.
\newblock \emph{Nature communications} 5.

\bibitem[Beuwer et~al.(2015)Beuwer, Prins, and Zijlstra]{beuwer2015stochastic}
Beuwer, M.~A., M.~W. Prins, and P.~Zijlstra, 2015.
\newblock Stochastic protein interactions monitored by hundreds of
  single-molecule plasmonic biosensors.
\newblock \emph{Nano letters} 15:3507--3511.

\bibitem[Sitters et~al.(2015)Sitters, Kamsma, Thalhammer, Ritsch-Marte,
  Peterman, and Wuite]{sitters2015acoustic}
Sitters, G., D.~Kamsma, G.~Thalhammer, M.~Ritsch-Marte, E.~J. Peterman, and
  G.~J. Wuite, 2015.
\newblock Acoustic force spectroscopy.
\newblock \emph{Nature methods} 12:47--50.

\bibitem[van Oijen(2011)]{VanOijen2011}
van Oijen, A.~M., 2011.
\newblock {Single-molecule approaches to characterizing kinetics of
  biomolecular interactions}.
\newblock \emph{Current Opinion in Biotechnology} 22:75--80.
\newblock
  \urlprefix\url{http://linkinghub.elsevier.com/retrieve/pii/S0958166910001904}.

\bibitem[Visser et~al.(2015)Visser, van IJzendoorn, and Prins]{Visser}
Visser, E.~W., L.~J. van IJzendoorn, and M.~W. Prins, 2015.
\newblock Particle Motion Analysis Reveals Nanoscale Bond Characteristics and
  Enhances Dynamic Range for Biosensing.
\newblock \emph{Submitted} .

\bibitem[Schafer et~al.(1991)Schafer, Gelles, Sheetz, and
  Landick]{DorothyA.SchaferJeffGellesMichaelP.Sheetz1991}
Schafer, D.~A., J.~Gelles, M.~P. Sheetz, and R.~Landick, 1991.
\newblock {Transcription by single molecules of RNA polymerase observed 'by
  light microscopy} .

\bibitem[Brinkers et~al.(2009)Brinkers, Dietrich, de~Groote, Young, and
  Rieger]{Brinkers2009}
Brinkers, S., H.~R. Dietrich, F.~H. de~Groote, I.~T. Young, and B.~Rieger,
  2009.
\newblock The persistence length of double stranded DNA determined using dark
  field tethered particle motion.
\newblock \emph{The Journal of chemical physics} 130:215105.

\bibitem[Milstein et~al.(2011)Milstein, Chen, and Meiners]{milstein2011bead}
Milstein, J., Y.~Chen, and J.-C. Meiners, 2011.
\newblock Bead size effects on protein-mediated DNA looping in
  tethered-particle motion experiments.
\newblock \emph{Biopolymers} 95:144--150.

\bibitem[Fan(2012)]{Fan2012}
Fan, H.-F., 2012.
\newblock Real-time single-molecule tethered particle motion experiments reveal
  the kinetics and mechanisms of Cre-mediated site-specific recombination.
\newblock \emph{Nucleic acids research} 40:6208--6222.

\bibitem[Kumar et~al.(2014)Kumar, Manzo, Zurla, Ucuncuoglu, Finzi, and
  Dunlap]{kumar2014enhanced}
Kumar, S., C.~Manzo, C.~Zurla, S.~Ucuncuoglu, L.~Finzi, and D.~Dunlap, 2014.
\newblock Enhanced tethered-particle motion analysis reveals viscous effects.
\newblock \emph{Biophysical journal} 106:399--409.

\bibitem[Scheepers(2015)]{Max}
Scheepers, M., 2015.
\newblock Time-dependent tethered particle motion for measuring dissociation
  kinetics of short complementary DNA oligonucleotides.
\newblock master thesis, Eindhoven University of Technology.

\bibitem[Plimpton(1995)]{plimpton1995fast}
Plimpton, S., 1995.
\newblock Fast parallel algorithms for short-range molecular dynamics.
\newblock \emph{Journal of computational physics} 117:1--19.

\bibitem[Zhang et~al.(2003)Zhang, Horsch, Lamm, and Glotzer]{Zhang2003}
Zhang, Z., M.~A. Horsch, M.~H. Lamm, and S.~C. Glotzer, 2003.
\newblock Tethered nano building blocks: Toward a conceptual framework for
  nanoparticle self-assembly.
\newblock \emph{Nano Letters} 3:1341--1346.

\bibitem[Naderi and van~der Schoot(2014)]{Naderi2014}
Naderi, S., and P.~van~der Schoot, 2014.
\newblock {Effect of bending flexibility on the phase behavior and dynamics of
  rods.}
\newblock \emph{The Journal of chemical physics} 141:124901.
\newblock \urlprefix\url{http://www.ncbi.nlm.nih.gov/pubmed/25273468}.

\bibitem[Underhill and Doyle(2004)]{Underhill2004}
Underhill, P.~T., and P.~S. Doyle, 2004.
\newblock {On the coarse-graining of polymers into bead-spring chains}.
\newblock \emph{Journal of Non-Newtonian Fluid Mechanics} 122:3--31.
\newblock
  \urlprefix\url{http://linkinghub.elsevier.com/retrieve/pii/S0377025704001909}.

\bibitem[Brenner(1961)]{Brenner1961}
Brenner, H., 1961.
\newblock The slow motion of a sphere through a viscous fluid towards a plane
  surface.
\newblock \emph{Chemical engineering science} 16:242--251.

\bibitem[Happel and Brenner(1983)]{happel1983low}
Happel, J., and H.~Brenner, 1983.
\newblock Low Reynolds number hydrodynamics: with special applications to
  particulate media, volume~1.
\newblock Springer Science \& Business Media.

\bibitem[Sch{\"a}ffer et~al.(2007)Sch{\"a}ffer, N{\o}rrelykke, and
  Howard]{Scha2007}
Sch{\"a}ffer, E., S.~F. N{\o}rrelykke, and J.~Howard, 2007.
\newblock Surface forces and drag coefficients of microspheres near a plane
  surface measured with optical tweezers.
\newblock \emph{Langmuir} 23:3654--3665.

\bibitem[Doi and Edwards(1988)]{doi1986theory}
Doi, M., and S.~F. Edwards, 1988.
\newblock The theory of polymer dynamics, volume~73.
\newblock oxford university press.

\bibitem[Phillips et~al.(2012)Phillips, Kondev, Theriot, and
  Garcia]{phillips2012physical}
Phillips, R., J.~Kondev, J.~Theriot, and H.~Garcia, 2012.
\newblock Physical biology of the cell.
\newblock Garland Science.

\bibitem[Merkus(2015)]{Koen}
Merkus, K., 2015.
\newblock Unraveling single-bond kinetics in tethered particle motion
  experiments using molecular dynamics simulations.
\newblock master thesis, Eindhoven University of Technology.

\end{thebibliography}

\clearpage
\newgeometry{left=20mm,bottom=1.6cm}
\section*{Figure Legends}
\begin{figure}[htb]
\includegraphics[width=\linewidth]{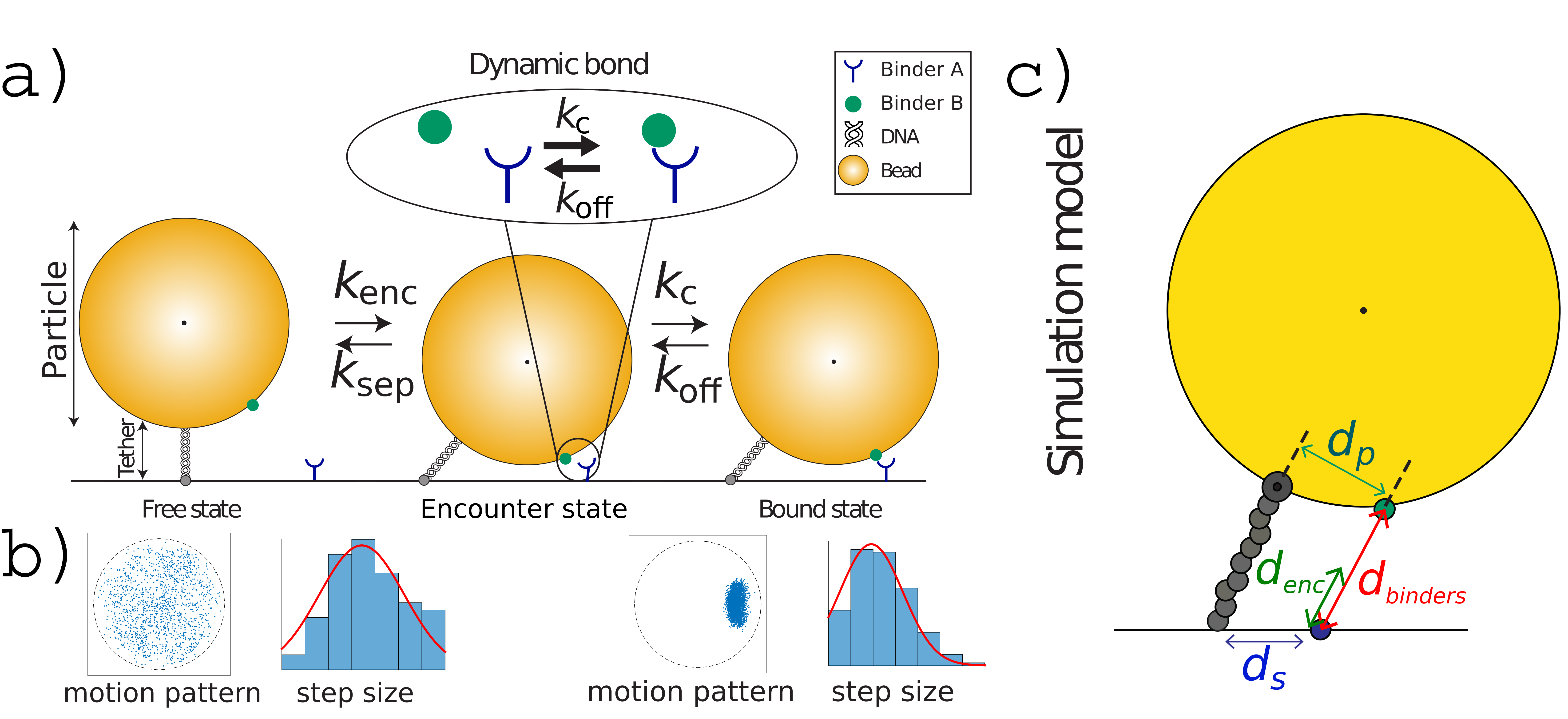}
\caption{Tethered Particle Motion with a secondary bond. a) The particle is bound to the substrate by a DNA tether. This figure provides an overview of the two-step process, in which the particle reversibly transfers between the \emph{free}, \emph{encounter} and \emph{bound} state. The first step is governed by the Brownian motion of the particle with encounter rate $k_{\rm enc}$ and separation rate $k_{\rm sep}$. The second step is governed by the binding process between molecule A and B, which are drawn generically in this figure. The molecular complexation and decomplexation rates are $k_{\rm c}$ and $k_{\rm off}$. b) A motion pattern of the particle in the free state (left) and the bound state (right). A motion pattern is the result of periodically imaging the particle and plotting all observed in-plane particle positions in one figure. The motion pattern of a free particle is fundamentally different from the motion pattern of a bound particle. The step size is the in-plane distance the particle travels between two frames. The step sizes are also affected by the binding process, the average step size is lower in the bound state. c) Schematic representation of the used simulation model. The tether is modeled as a string of beads, there is a binding patch on the particle at a perpendicular distance $d_p$ from the tether point and a binding patch on the substrate at a distance $d_s$ from the origin. The encounter distance is denoted by $d_{\rm enc} $ and the distance between the two binding patches by $d_{\rm binders}$. We refer to the encounter state when $d_{\rm binders} < d_{\rm enc}$. The dashed curves indicate the interaction range for both binding patches: when the curves overlap the particle is in the encounter state.}
\label{fig:fig1}
\end{figure}

\subsubsection*{Figure \ref{fig:fig1}(a,b,c)}
\begin{figure}[htb]
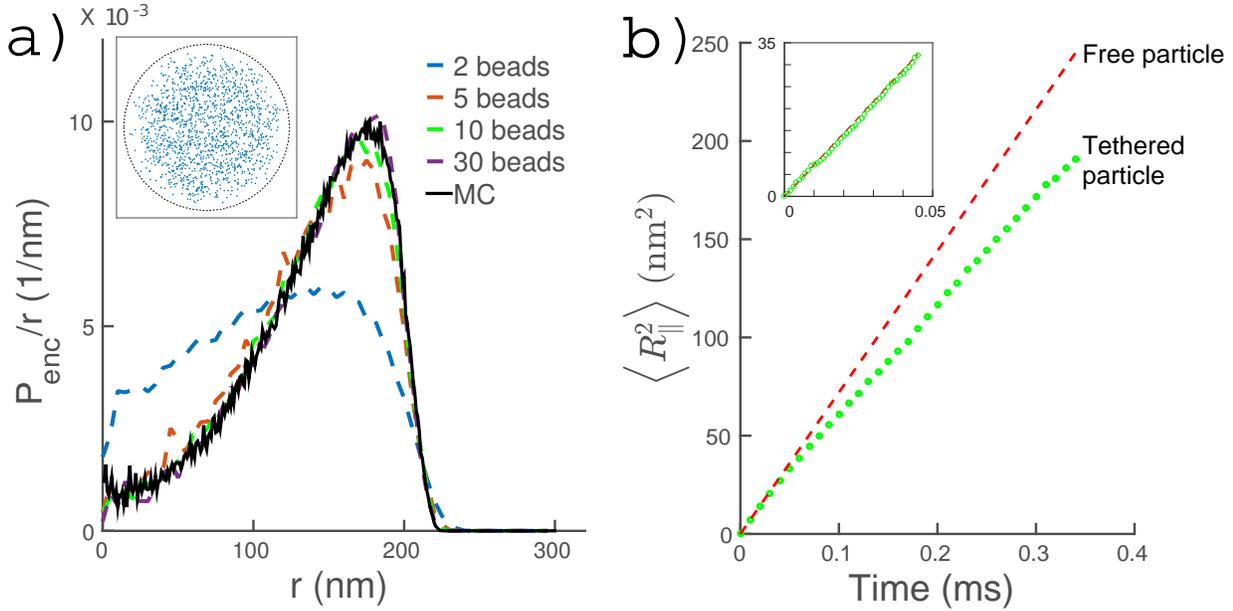

\includegraphics[width=0.5\linewidth]{number_beads_dashed_inset.pdf}
\includegraphics[width=0.5\linewidth]{TimeBenchmark2.pdf}
\caption{a) The encounter distribution per in-plane radius $P_{\rm enc}/r$ as function of the in-plane radius for a particle with diameter 1 $\mu$m and a tether with length 50 nm. The black line represents results obtained by Monte Carlo simulations with a chain of 50 segments. The dashed lines represent results from MD simulations. The states in which the distance between the edge of the bead and the substrate is smaller than 10 nm are defined as encounter states. It can be observed that the MD results converge to the MC results for an increasing number of beads on a string. The inset shows the motion pattern resulting from the MD simulation with 10 beads. The encounter distribution is obtained from these data points in combination with the $z$-coordinate of the particle. b) Comparison of the in-plane mean-squared displacement of the tethered particle and the mean squared displacement of a free particle starting at the same position. For small times the mean-squared displacement develops equally for a tethered and a free particle, but for larger times the displacement of the particle is smaller due to its confinement and the fact that it is on average closer to the substrate. Green dots are obtained by averaging over 3000 simulations.}
\label{fig:benchmark}
\end{figure}

\clearpage
\subsubsection*{Figure \ref{fig:benchmark}(a,b)}
\begin{figure}[htb]
\includegraphics[width=\linewidth]{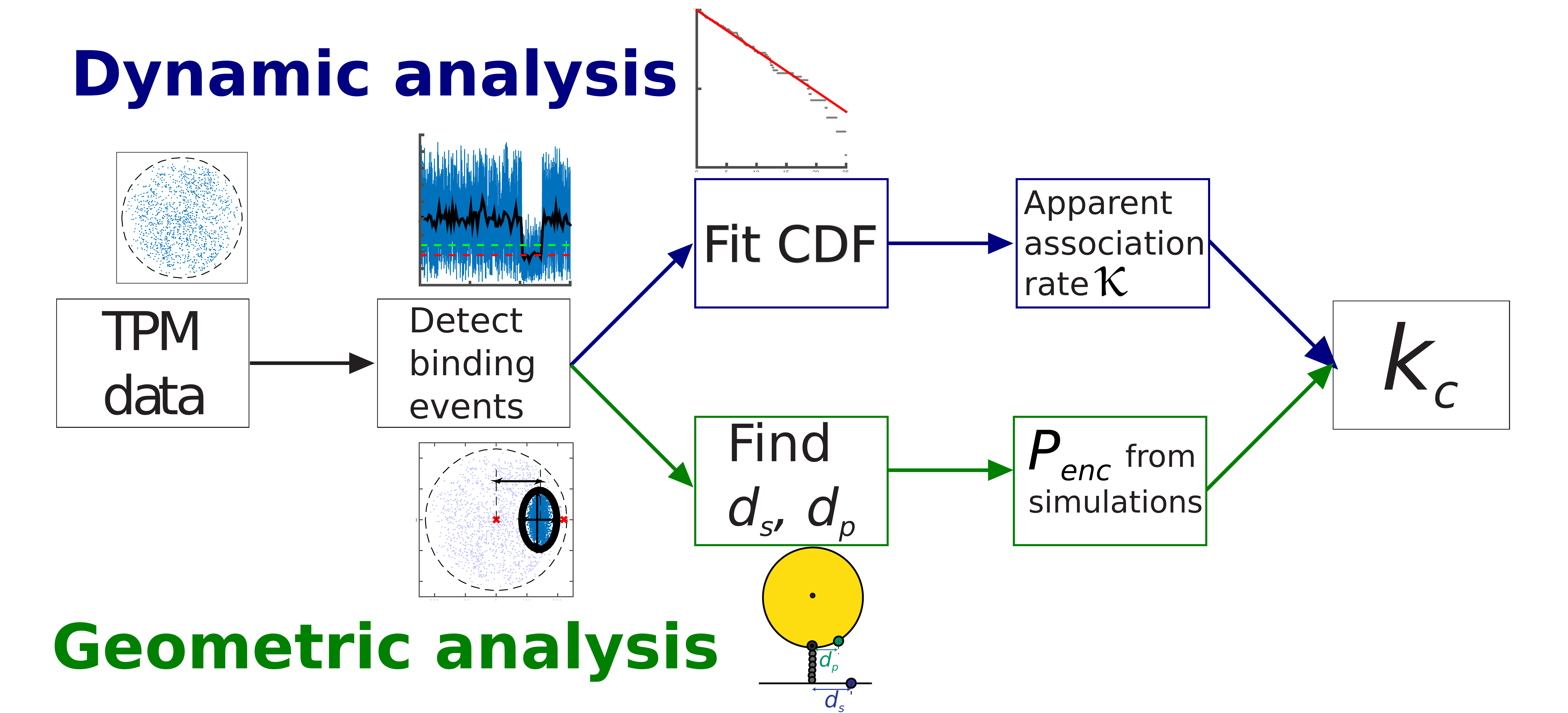}
\caption{A schematic representation of the data processing algorithm. The total analysis can be subdivided in dynamic analysis and geometric analysis. The dynamic analysis involves tracking the time between subsequent bonds to end up with an apparent association rate $\kappa$. The geometric analysis involves the characterization of a bound motion pattern, relating this to the most probable value of $d_s$ and $d_p$ and find the corresponding $P_{\rm enc}$. Finally, both analyses are merged to find our molecular property: the complexation rate $k_{\rm c}$ }
\label{fig:scheme}
\end{figure}

\clearpage
\subsubsection*{Figure \ref{fig:scheme}}
\begin{figure}[htb]
\includegraphics[width=0.42\linewidth]{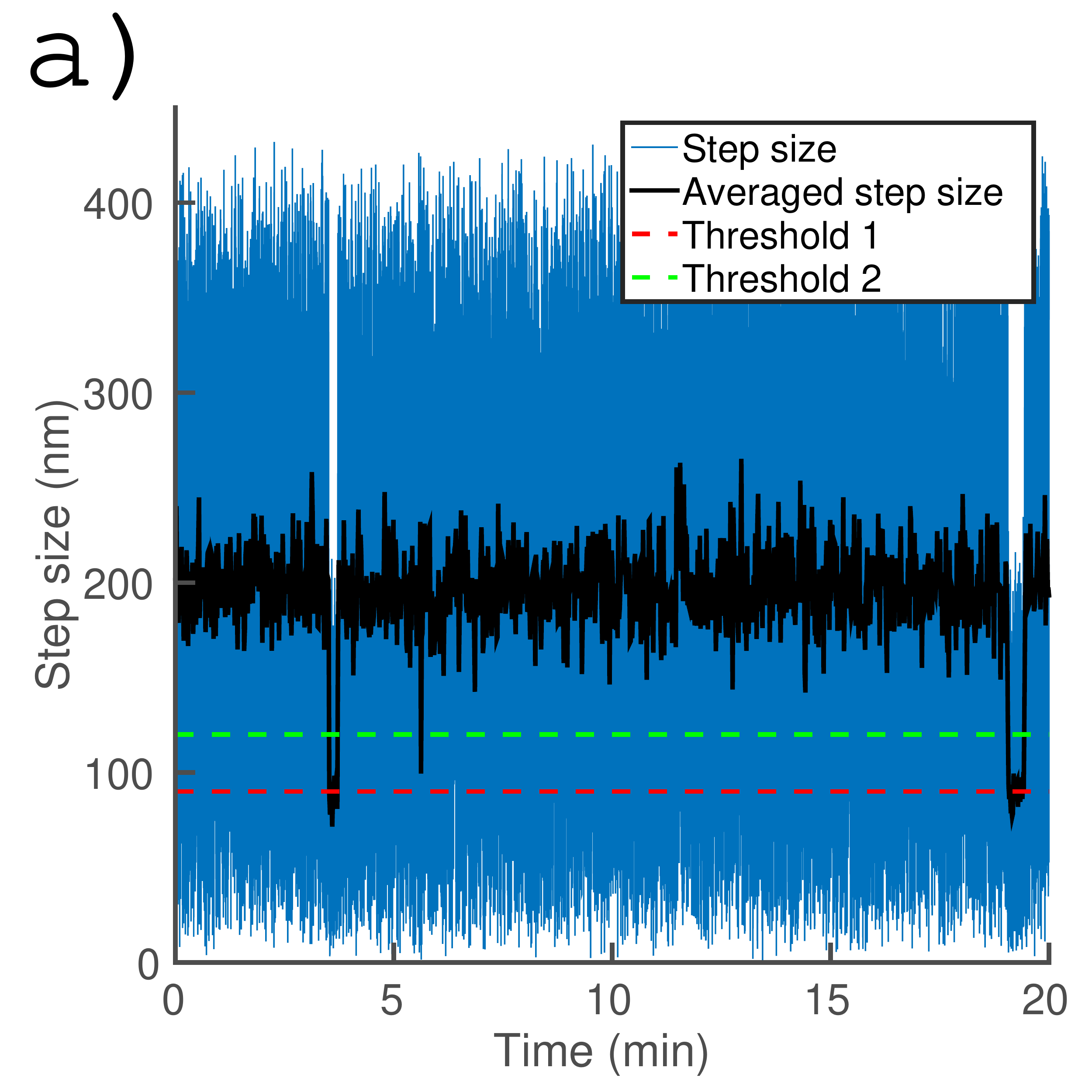}
\includegraphics[width=0.42\linewidth]{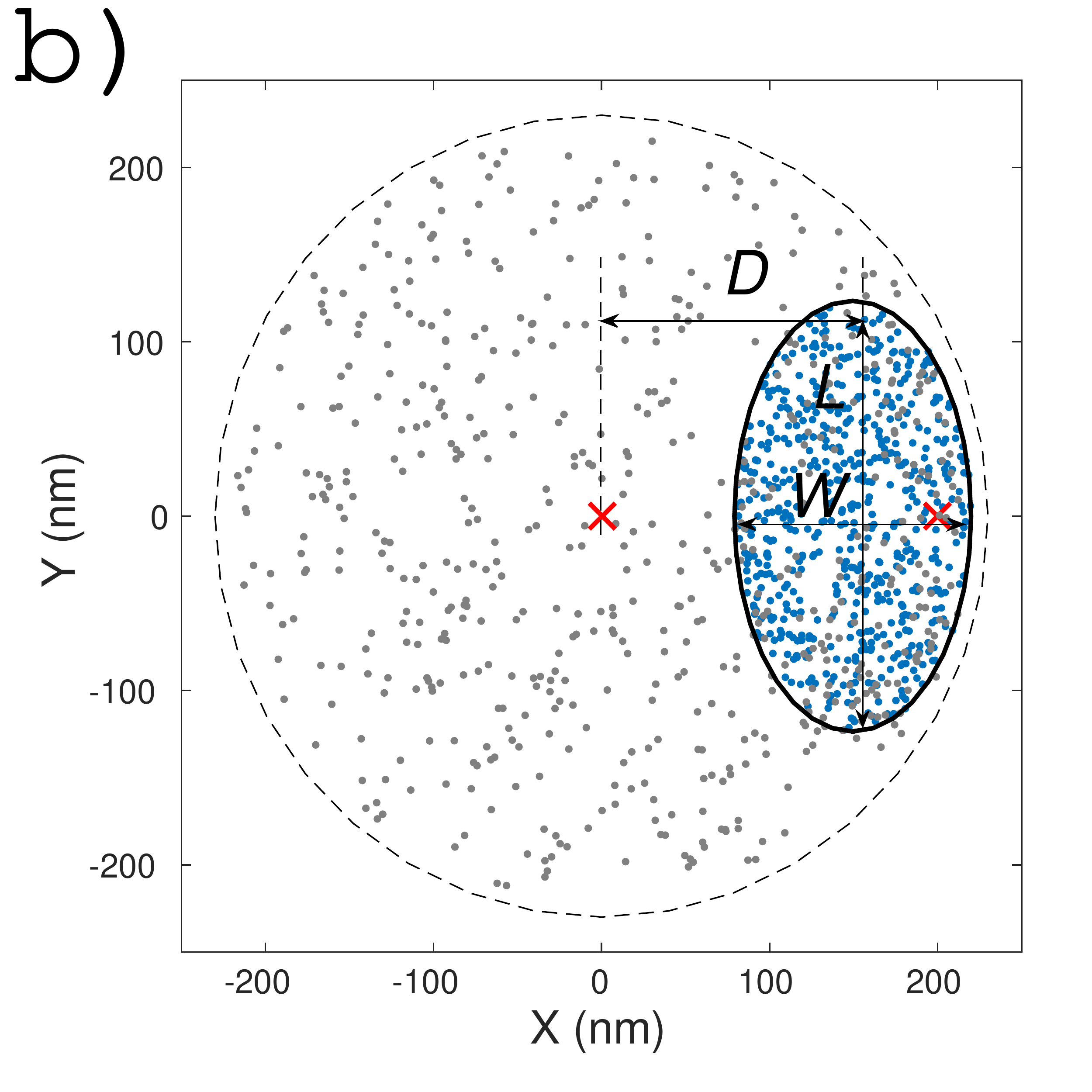}
\includegraphics[width=0.42\linewidth]{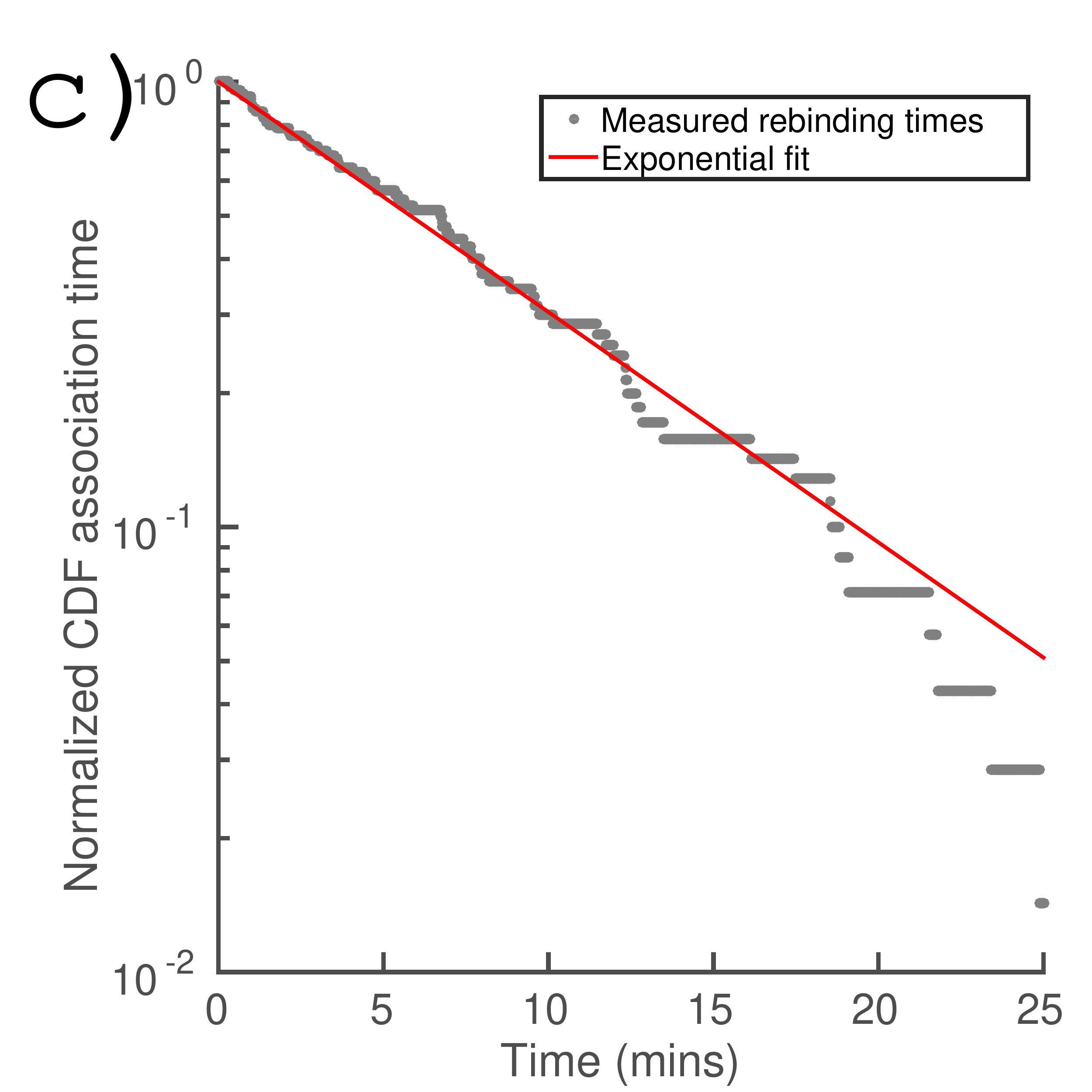}
\caption{Application of the data processing algorithm of Fig.~\ref{fig:scheme} to data generated by a mock model. a) The step sizes obtained from the mock model. Indicated are the step size (in blue), the step size averaged over 30 frames (in black), threshold 1 (in red), and threshold 2 (in green). This allows for detection of binding events following the procedure developed by~\cite{Max} . b) Motion pattern from the mock model. The blue points are generated in the `bound' state and the grey points belonging in the `free' or `encounter' state. Three characteristic geometric parameters may be extracted from the blue points: the length $L$, the width $W$ and the average distance to the central axis $D$. This set of parameters can be linked to the most probable positions of the binders $d_s$ and $d_p$. c) An exponential fit of the cumulative distribution function (CDF) of association times. The vertical axis scales logarithmically. The grey dots represent `measured' rebinding times from our mock model and the red line is an exponential fit of the form $P(t) = P_0 \exp(-\kappa t)$ with $P_0$ is 1, which yielded $\kappa=(1.7 \pm 0.3) \cdot 10^{-3}$ s$^{-1}$ (mean $\pm$ standard deviation).}
\label{fig:analysis}
\end{figure}

\clearpage
\newgeometry{left=20mm,bottom=0.1cm, top=15mm}
\subsubsection*{Figure \ref{fig:analysis}(a,b,c)}
\clearpage
\clearpage

\newgeometry{left=35mm,bottom=3.0cm}
\beginsupplement

\section{Implementing hydrodynamic effects near a surface}
As mentioned is section~\ref{sec:drag_bead}, the drag on the bead near a surface is no longer the isotropic drag, but the drag increases in a distinct matter for the parallel and perpendicular component. In particular, we use the relations
\begin{align}
\gamma_{\parallel} &= \frac{\gamma_0}{1-\frac{9}{16}q + \frac{1}{8}q^3-\frac{45}{256}q^4-\frac{1}{16}q^5},\\
\gamma_{\bot} &=  \frac{\gamma_0}{1-\frac{9}{8}q + \frac{1}{2}q^3-\frac{57}{100}q^4+\frac{1}{5}q^5},
\end{align}
for the parallel drag $\gamma_{\parallel}$ and the perpendicular drag $\gamma_{\bot}$ for a particle near a wall at $z=0$. Here $q=R/z$, $\eta$ is the dynamic viscosity of the solvent and $\gamma_0 = 6 \pi \eta R$, the Stokes drag in an unbounded liquid. \\[2mm]
We define $c_{\parallel}$ and $c_{\bot}$
\begin{align}
c_{\parallel} &= \frac{1}{1-\frac{9}{16}q + \frac{1}{8}q^3-\frac{45}{256}q^4-\frac{1}{16}q^5},\\
c_{\bot} &=  \frac{1}{1-\frac{9}{8}q + \frac{1}{2}q^3-\frac{57}{100}q^4+\frac{1}{5}q^5},
\end{align}
so that $\gamma_{\parallel} = c_{\parallel} \gamma_0$ and $\gamma_{\bot} = c_{\bot} \gamma_0$. \\[2mm]
In the LAMMPS code~\cite{plimpton1995fast} this can be implemented by adjusting the random force and the drag force in the file fix\_langevin.cpp.

\begin{verbatim}
  //For loops loops over all atoms in bin
  //and applies random force, after that
  //drag force is applied.
  for (int i = 0; i < nlocal; i++) {
  //Assign the perpendicular and parrallel
  //component a value depending on the z-coordinate
  zs = R_bead/(R_bead+x[i][2]);
  //This is a rescaled version of z, 
  //convenient for the expressions of the parr and perp drag
  //Use Faxen's law and an interpolation formula given by Schaffer et al.
  //to determine both coeffs
  cparr = 1/(1-0.5625*zs+0.125*pow(zs,3)-0.175781*pow(zs,4)-0.0625*pow(zs,5));
  cperp=1/(1-1.125*zs+0.5*pow(zs,3)-0.57*pow(zs,4)+0.2*pow(zs,5));
    if (mask[i] & groupbit) {
      if (Tp_TSTYLEATOM) tsqrt = sqrt(tforce[i]);
      if (Tp_RMASS) {
        gamma1 = -rmass[i] / t_period / ftm2v;
        gamma2 = sqrt(rmass[i]) * sqrt(24.0*boltz/t_period/dt/mvv2e) / ftm2v;
        gamma1 *= 1.0/ratio[type[i]];
        gamma2 *= 1.0/sqrt(ratio[type[i]]) * tsqrt;
      } else {
        gamma1 = gfactor1[type[i]];
        gamma2 = gfactor2[type[i]] * tsqrt;
      }

      fran[0] = sqrt(cparr)*gamma2*(random->uniform()-0.5);
      fran[1] = sqrt(cparr)*gamma2*(random->uniform()-0.5);
      fran[2] = sqrt(cperp)*gamma2*(random->uniform()-0.5);

     fdrag[0] = cparr*gamma1*v[i][0];
     fdrag[1] = cparr*gamma1*v[i][1];
     fdrag[2] = cperp*gamma1*v[i][2];

\end{verbatim}

\noindent
If the desired coordinate for the wall is not the z-component, this can be incorporated by requiring the arguments `face' and `coord' to describe the location of the wall and process this by
\begin{verbatim}
  //Check if the wall face is set in a valid way
  if (strcmp(arg[iarg],"xlo") == 0) wallface = XLO;
  else if (strcmp(arg[iarg],"xhi") == 0) wallface = XHI;
  else if (strcmp(arg[iarg],"ylo") == 0) wallface = YLO;
  else if (strcmp(arg[iarg],"yhi") == 0) wallface = YHI;
  else if (strcmp(arg[iarg],"zlo") == 0) wallface = ZLO;
  else if (strcmp(arg[iarg],"zhi") == 0) wallface = ZHI;
  else error->all(FLERR, "Wall face should be a coordinate with 'lo' or 'hi'");
  dim = wallface / 2; //0 for x, 1 for y, 2 for z
  side = wallface % 2; //0 for lo, 1 for hi
\end{verbatim}
And then apply them to the component in the relevant direction

\begin{verbatim}
//Assign the perpendicular and parrallel 
//component a value depending on the coordinate
//in the direction of the wall 
//This is a rescaled version of the coordinate, 
//convenient for the expressions of the parr and perp drag
if (!side) xs = R_bead/(x[i][dim]-coord0); 
else      xs = R_bead/(coord0 - x[i][dim]);
if (xs<0.0) error->all(FLERR, "Rescaled component is negative");
if (xs>1.0) error->all(FLERR, "Bead is intersecting wall");

//Use Faxen's law and an interpolation formula given by
//Schaffer et al. to determine both coeffs
cparr = 1.0/(1.0-0.5625*xs+0.125*pow(xs,3)-0.175781*pow(xs,4)-0.0625*pow(xs,5));
cperp=1.0/(1.0-1.125*xs+0.5*pow(xs,3)-0.57*pow(xs,4)+0.2*pow(xs,5));
if (dim ==0){
        cx = cperp;
        cy = cparr;
        cz = cparr;
      }
      else if (dim==1){
        cx = cparr;
        cy = cperp;
        cz = cparr;
      }
      else if (dim==2){
        cx = cparr;
        cy = cparr;
        cz = cperp;
      }

\end{verbatim}

\clearpage

\section{Optimizing engineering parameters}
In designing a TPM system it is worthwhile to investigate the influence of several engineering parameters to maximize the number of potential binding events. In this respect, the tether length $l$ and the bead diameter $R$ are relevant parameters, but the used frame rate may aso be of significant influence. We present a way to optimize such experiments.\\[2mm]
For a maximum number of binding events to occur, the reactive surface on the substrate and the reactive surface on the particle should be within encounter distance as much as possible. We aim to define a parameter that quantifies the amount of surface on the particle and the substrate being within encounter distance. Since the system is axisymmetric, we may define a contact parameter $\zeta$ as
\begin{equation}
\zeta = \int\displaylimits_{A_{s}} \int\displaylimits_{A_{p}} P_{\rm enc}(d_p,d_s) \mathrm{d}A_{ring,p} \mathrm{d}A_{ring,s}, \label{eq:int_zeta}
\end{equation}
where $P_{\rm enc}$ represents the probability that a binding spot on the particle and a binding spot on the substrate are within interaction range. $\mathrm{d}A_{ring,p}$ and $\mathrm{d}A_{ring,s}$ are ring-shapen area elements on the particle and the substrate over which we integrate.
\\[2mm]
For a maximum number of potential binding events, $\zeta$ should be maximized. For fixed bead diameter of 1 $\mu$m and fixed frame rate of 30 Hz, we have calculated $\zeta$ for several tether lengths $L$ and this lead to the trivial optimal tether length $L=0$. In other words, with a longer tether the particle will encounter the substrate less often and the extra available potential encounter positions are not able to make up for this.
\\[2mm]
However, in designing the optimal measurement set-up, the contact area is not the only important parameter. Namely, not every secondary bond is detectable and we want to optimize the number of \emph{detectable} bonds. The specific position of the binding molecules determines the step size $R_{step} = |\vec R(t+\Delta t) - \vec R(t)|$, the in-plane distance that the particle travels between two frames, which is used to detect secondary bonds. In order to encompass this effect and include only the potential detectable bonds, we define a refined contact parameter
\begin{equation}
\xi = \int\displaylimits_{d_{s,min}}^{d_{s,max}} \int\displaylimits_{A_{p}} P_{\rm enc}(d_p,d_s) \mathrm{d}A_{ring,p} \mathrm{d}A_{ring,s}, \label{eq:int_zeta2}
\end{equation}
where $d_{s,min}$ is the minimum $d_s$ that leads to a detectable bond and $d_{s,max}$ is the maximum possible $d_s$. Our preliminary results delivered an optimal tether length of 60 nm for a bead diameter of 1 $\mu$m and frame rate 30 Hz~\cite{Koen}.

\end{document}